\documentclass{iaus}
\usepackage{graphicx}
\usepackage{sidecap}

\title[Oscillations, granulation phenomena and sun-like spots] 
{Stellar noise and planet detection. \\
{\Large I. Oscillations, granulation and sun-like spots}}

\author[X. Dumusque et al.]   
{X. Dumusque$^{1,}$$^2$,
N. C. Santos$^2$,
S. Udry$^1$,
C. Lovis$^1$
\and X. Bonfils$^3$}

\affiliation{$^1$Observatoire de Gen\`eve, 51 ch. des Maillettes, 1290 Sauverny, Switzerland \\[\affilskip]
$^2$Centro de Astrof{\'\i}sica, Universidade do Porto, Rua das Estrelas, 4150-762 Porto, Portugal \\[\affilskip]
$^3$Universit\'e J. Fourier (Grenoble 1)/CNRS, Laboratoire d'Astrophysique de Grenoble, France}

\pubyear{2010}
\volume{276}  
\pagerange{0--1}
\jname{The Astrophysics of Planetary Systems: Formation, \\Structure, and Dynamical Evolution}
\editors{Alessandro Sozzetti, Mario G. Lattanzi \& Alan P. Boss, eds.}

\begin{document}

\maketitle

\begin{abstract}
Spectrographs like HARPS can now reach a sub-ms$^{-1}$ precision in radial-velocity (RV)
(\cite[Pepe \& Lovis (2008)]{Pepe-2008}). At this level of accuracy, we start 
to be confronted with stellar noise produced by 3 different physical phenomena: oscillations, 
granulation phenomena (granulation, meso- and super-granulation) and activity.
On solar type stars, these 3 types of perturbation can induce ms$^{-1}$ RV variation, but on different time scales:
 3 to 15 minutes for oscillations, 15 minutes to 1.5 days for granulation phenomena and 10 to 50 days for activity.
The high precision observational strategy used on HARPS, 1 measure
per night of 15 minutes, on 10 consecutive days each month, is optimized, due to a long exposure time, to average
out the noise coming from oscillations (\cite[Dumusque \etal\ (2010a)]{Dumusque-2010a})
 but not to reduce the noise coming from granulation and activity (\cite[Dumusque \etal\ (2010a)]{Dumusque-2010a} 
 and \cite[Dumusque \etal\ (2010b)]{Dumusque-2010b}). The smallest planets found with this strategy (\cite[Mayor \etal\ (2009)]{Mayor-2009})
seems to be at the limit of the actual observational strategy and not at the limit of the instrumental
precision. To be able to find Earth mass planets in the habitable zone of solar-type stars (200 days for a K0 dwarf), new observational strategies, averaging out simultaneously all type of stellar noise, are required.
\\
\end{abstract}
\vspace{0.1cm}


\noindent
\underline{Generating synthetic radial-velocity measurements with stellar noise}

\vspace{0.1cm}

To simulate the effect of new observational strategies, we first have to generate synthetic RV measurements that contain all the considered type of noises: oscillation, granulation phenomena and short-term activity. 

For the two first perturbations, which have a typical time scale less than a few days, we use asteroseismology measurements to derive the corresponding noise levels. The technique used consists in calculating the velocity power spectrum density (VPSD) and in fitting it with a function that depend on the different type of noises considered
%
Once the fit is done, we can generate synthetic RVs containing all the considered noises by calculating the inverse fourrier transform of the fitted function (see \cite[Dumusque \etal\ (2010a)]{Dumusque-2010a} for more details)

Short-term activity noise induced by sun-like spots have a timescale of 10 to 50 days. Asteroseismology measurements, spanning a maximum of 10 days, are too short to address this type of noise whereas longer observations are not precise enough. We therefore need to simulate the RV effect of spots using Sun observations. Starting with simple sunspot properties, such as their number, their position in latitude and longitude, their lifetime, we generate a realistic model of sun-activity using a Poisson process for spot appearance. The RV effect induced by these spots is then calculated using the program SOAP (X. Bonfils \& N.C. Santos (2010), in prep.), which calculate the RV effect of one spot given its size and position (see \cite[Dumusque \etal\ (2010b)]{Dumusque-2010a} for more details).

\vspace{0.5cm}

\noindent
\underline{An efficient and affordable observational strategy}
\vspace{0.1cm}

\begin{figure}
\centering
\includegraphics[width=7.6cm]{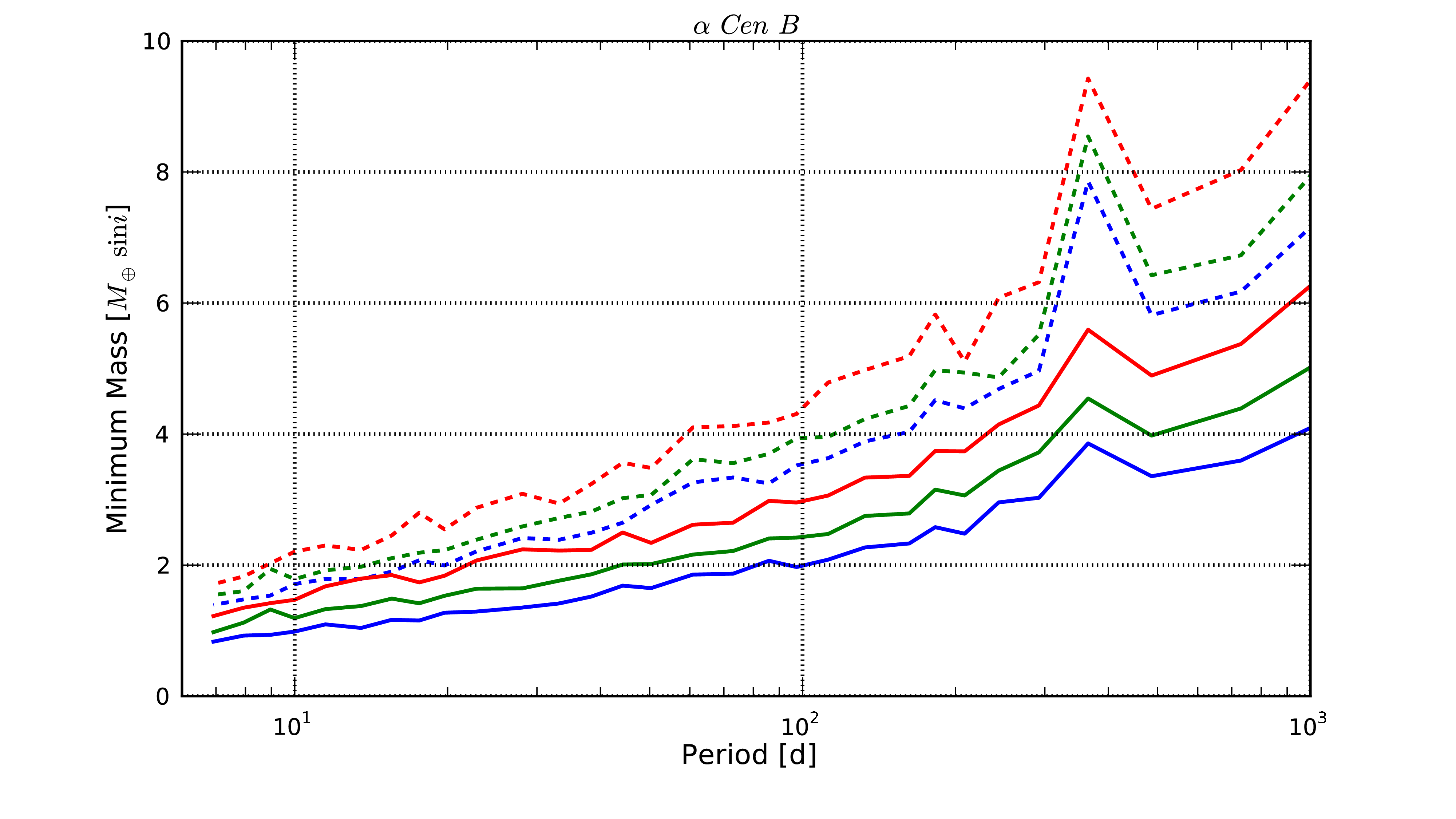}
\caption{Mass-period detection limits for different strategies (continuous lines: 3N strategy and dashed lines: 1N strategy) and for different activity levels ($\log{(R'_{HK})}=-5$ (blue), $-4.9$ (green) and $-4.8$ (red) from bottom to top for each strategy).}
\label{fig:1}
\end{figure}

\begin{figure}
\begin{center}
\includegraphics[width=6cm]{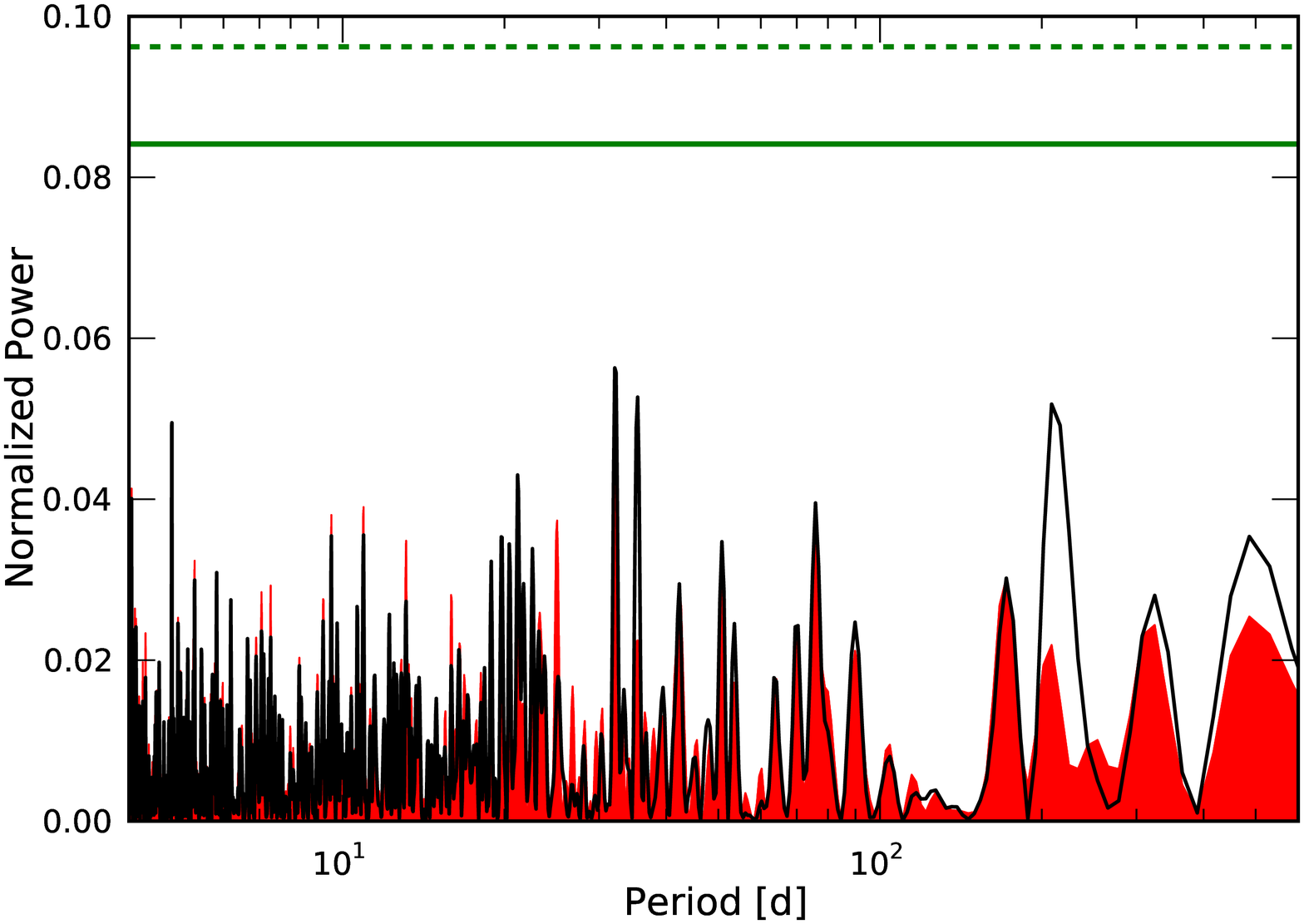}
\includegraphics[width=6cm]{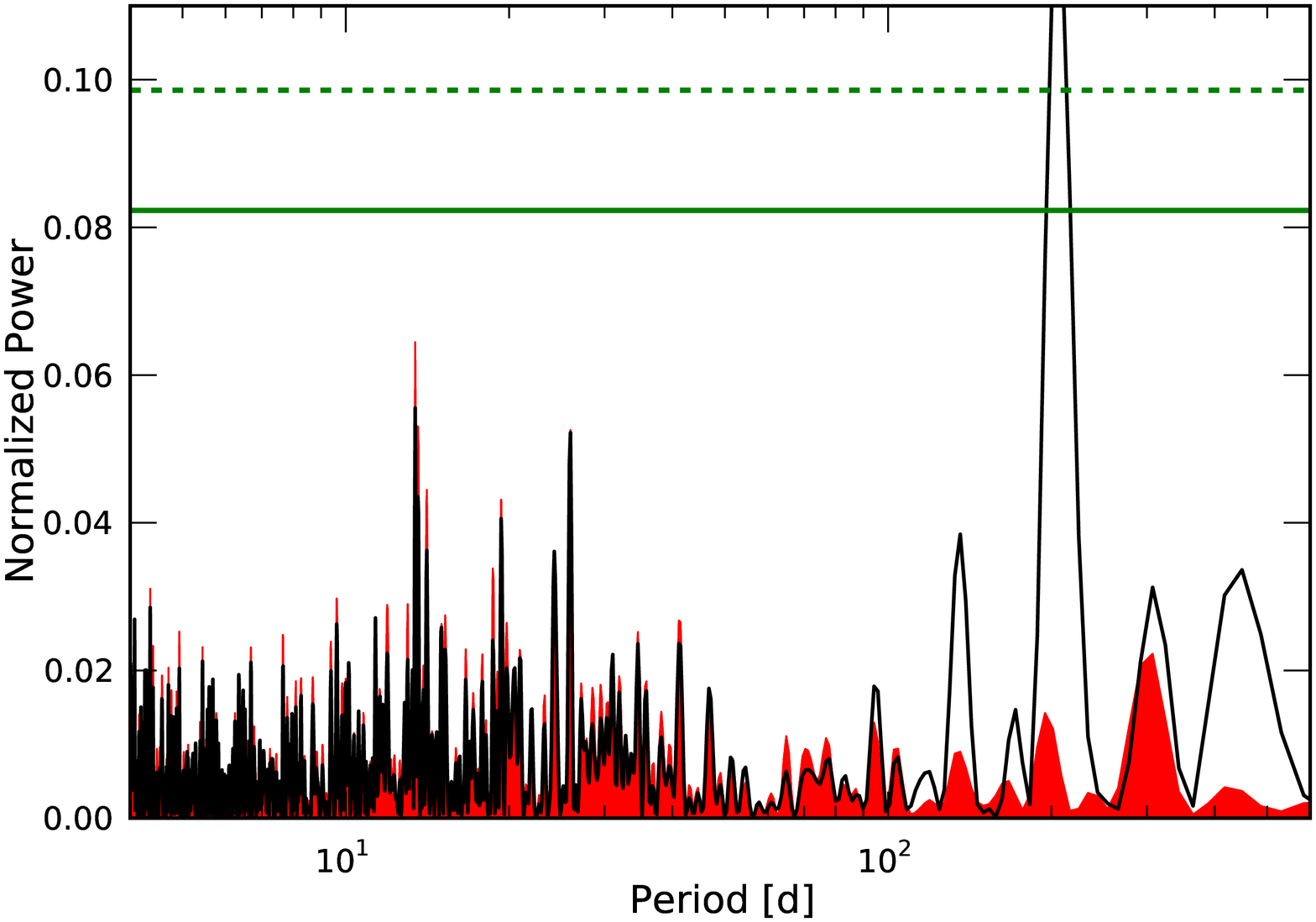}
\caption{Detection simulation of a 2.5\,$M_{\oplus}$ habitable planet orbiting an early K dwarf (200 days of period), with a star activity level fixed at $\log{(R'_{HK})}=-4.9$. \emph{Left :} Periodogram for the 1N strategy with the 1\,\% FAP (continuous line) and the 0.1\,\% FAP (dashed line). The red (gray) periodogram corresponds to the noise, and the black, to noise\,$+$\,planet. \emph{Right :} Same but for the 3N strategy.}
\label{fig:2}
\end{center}
\end{figure}

Once synthetic RV measurements containing oscillation, granulation phenomena and activity noises are generated, we choose an observational calendar (10 nights per month, 8 months a year on 4 years) and calculate for different observational strategies the planetary mass detection limits using 1\,\% False Alarm Probability (FAP) in periodograms (see Fig. \ref{fig:1}). Depending on the activity level considered, the best tested strategy, 3 measurements per night of 10 minutes each, each 3 nights (3N strategy), could detect planet of 2.5 to 3.5\,$M_{\oplus}$ in the habitable region of early K dwarfs (200 days of period). This strategy gives detection limit 40\,\% better than the present strategy used on HARPS (1N strategy). In Fig. \ref{fig:2}, we can see the detection simulation of a 2.5 $M_{\oplus}$ habitable planet orbiting an early K dwarf. In this case, the best tested strategy is able to resolve the planet, whereas the present HARPS strategy is not (see \cite[Dumusque \etal\ (2010b)]{Dumusque-2010a} for more details).

\end{document}